\documentstyle[aps,twocolumn]{revtex}
\begin{document}

\title{ Spectral measurement of the Hall angle response in normal state
cuprate superconductors}

\author{
M. Grayson$^{1,2,*}$, L. B. Rigal$^{1,\dag}$, D. C. Schmadel$^1$,
H. D. Drew$^{1,3}$, P.-J. Kung$^4$}

\address{
$^1$Department of Physics, University of Maryland, College Park, MD
20742\\
$^2$Walter Schottky Institut, Technische Universit\"at M\"unchen, D-85748 
Garching, Germany\\
$^3$Center for Superconductivity Research, University of Maryland, College 
Park, MD 20742\\
$^4$Advanced Fuel Research, Inc., East Hartford, CT}

\date{29 May 2002}

\maketitle
 
\begin{abstract}

We measure the temperature and frequency dependence of the complex Hall
angle for normal state YBa$_2$Cu$_3$O$_7$ films from dc to far-infrared
frequencies (20-250 cm$^{-1}$) using a new modulated polarization
technique.  We determine that the functional dependence of the Hall angle
on scattering does not fit the expected Lorentzian response.  We find
spectral evidence supporting models of the Hall effect where the
scattering $\Gamma_H$ is linear in T, suggesting that a single relaxation
rate, linear in temperature, governs transport in the cuprates.

\end{abstract}
\narrowtext

The normal state Hall effect in cuprate superconductors exhibits an
anomalous temperature dependence that cannot be explained using
conventional transport theory for metals.  According to the simple Drude
model, the resistivity of a metal and the cotangent of its Hall angle
$cot(\theta_H)=\frac{\sigma_{xx}}{\sigma_{xy}}$, should share the same
temperature dependence, both proportional to the scattering rate of the
charge carriers.  However, the normal state resistance of cuprate
superconductors is linear with temperature, $\rho \sim T$, while the Hall
angle has a robust $cot(\theta_H) \sim T^2$ behavior \cite{Chien1} over a
wide range of oxygen doping \cite{Harris}, and with substitutional doping
\cite{Chien2} in a variety of the cuprates \cite{Kendziora}.  This
apparent duality of scattering rates characterizes the anomalous Hall
transport in the cuprates.  Several theories approached the problem
assuming that two scattering rates were in fact involved, beginning with
the spin-charge separation model of Anderson wherein the two species of
quasiparticles each relaxed at the different observed
rates.\cite{Anderson} Subsequent explanations focused either on
alternative non-Fermi liquid mechanisms \cite{Coleman,Lee} or on the
effects of $k$-space scattering anisotropies
\cite{Ioffe,Zheleznyak,Stojkovic,Kotliar}.  The common feature of all the
above theories is a dominant term that is linear in the scattering rate,
$cot(\theta_H) \sim \gamma_H$.  In contrast, a recent theory by Varma and
Abrahams \cite{Varma} treats anisotropic scattering in a marginal
Fermi-liquid and predicts a {\em square}-scattering response,
$cot(\theta_H) \sim \gamma_H^2$.

These different models can be distinguished at finite frequency.  The
linear- and square-scattering models correspond to Lorentzian and
square-Lorentzian spectral responses respectively, and although Hall
experiments have been performed at finite frequencies
\cite{Kaplan,Parks,Cerne}, this paper is the first to study both
temperature {\em and} frequency dependence of the Hall response in a
frequency range that discerns a lineshape and extrapolates to the dc
limit.

We begin by reviewing the concept of a frequency dependent Hall angle
\cite{Drew} using the Drude model as an example of a Lorentzian response.  
All parameters are implicitly spectral, i.e. $\theta_H =
\theta_H(\omega)$, and in the present case of strong scattering,
$tan(\theta) \simeq \theta << 1$.  Quasiparticles circling at the
cyclotron frequency $\omega_H^*=eB/mc$ traverse a fraction
$\omega_H^*\tau_H^*$ of a cyclotron orbit during the time $\tau_H^*$
between scattering events.  The dc longitudinal current $j_x$ is thereby
deflected into $j_y$ by this small arc angle $\theta_H \simeq
\frac{j_y}{j_x} = \omega_H^* \tau_H^* = \omega_H^*/\gamma_H^*$, where
$\gamma_H^*$ is defined as the Hall scattering rate.  For the ac response,
we substitute $\gamma_H^* \rightarrow \gamma_H^* - i \omega$ yielding a
Lorentzian:

\begin{equation}
\theta_H(\omega) = \frac{\omega_H^*} {\gamma_H^* - i \omega}
\label{linscat}
\end{equation}

\noindent Linear scattering models of the anomalous Hall transport predict
this same Lorentzian form for the frequency response near the dc limit.

For the experiment at infrared frequencies, the Hall angle $\theta_H$
cannot be measured directly but must be deduced from transmission studies
of polarized light \cite{Kaplan,Parks,Cerne}.  The measurable quantity,
the Faraday angle $\theta_F$, is the angle of rotation of polarized light
induced after passing through a thin conducting film in the presence of a
normal magnetic field $B$.  In the thin film limit ($d << \lambda,
\delta$) with $d$ the film thickness, $\lambda$ the wavelength, and
$\delta$ the penetration depth, $\theta_H$ follows from $\theta_F$
according to Maxwell's equations:

\begin{equation}
\theta_H = \left( \frac{1+n}{Z_o \sigma_{xx}} + 1 \right) \theta_F
\label{theta_H}
\end{equation}

\noindent $\sigma_{xx}$ is the experimentally determined complex sheet
conductivity, $Z_o$ the free space impedance, and $n=3.4$ the refractive
index of the Si substrate.  With our highly conductive films the term in
parenthesis is near unity, so the functional dependence on $\sigma_{xx}$
is minimal.  $\theta_F$ and $\theta_H$ are both causal response functions,
so their Re and Im parts obey Kramers-Kronig relations and correspond to
real space rotation and ellipticity, respectively \cite{Drew}.

The cuprate sample investigated was a pulsed-laser deposited, twinned film
of 500~\AA~ YBa$_2$Cu$_3$O$_7$ on a 1$\times$1~cm$^2$ $\times$ 350~$\mu$m
insulating silicon substrate, with an intermediate strain relieving layer
of 100~\AA~ yttrium stabilized zirconate \cite{Kaplan}.  Partial data on a
second sample confirmed the results reported here.  The sample was mounted
in an 8~T Oxford Spectromag cryostat with the $B$-field oriented along the
optical axis, normal to the sample surface.  Two 8 cm diameter Kapton
cryostat windows on either side of the sample allowed direct optical
access.  Sourced by a broadband quartz Hg arc lamp, the spectra were
measured using a step-scan Fourier transform interferometer with wire-grid
polarized beam-splitters having a density of 40 wires per mm.  The novel
technique consisted of a mechanically rotating optical element which
modulated the polarization of light incident on the sample, and the
transmitted signal was measured with a bolometer detector using standard
lockin techniques at harmonics of the rotator frequency.  Measurements of
cyclotron resonance in GaAs verified the experimental technique.

Schematics for measuring Re($\theta_F$) and Im$(\theta_F)$ are shown in
the insets of Fig. \ref{Fig1}.  To measure the Faraday angle
Re($\theta_F$), polarized light was projected through a mechanically
rotating linear polarizer with rotation angle $\phi(t) = 2\pi f_{rot} t$,
($f_{rot} = $70 Hz).  The light then passed through the sample, striking a
stationary polarizer before reaching the bolometer detector.  The phase
shift in the second harmonic of the bolometer signal is identically the
Faraday rotation, Re($\theta_F$) =Re ($t_{xy}/t_{xx}$), where $t_{ij}$ is
the transmittance tensor relating the transmitted field in the $j$
direction to the incident field in the $i$ direction \cite{Cerne}.

Im($\theta_F$) is measured analogously (inset Fig.~\ref{Fig1}), with the
light modulated into right and left elliptical polarizations by a
mechanically rotating wave plate and by omitting the polarizer previously
in front of the detector.  The waveplate shifts the optical phase of the
field component along its extraordinary axis by the retardance
$\beta(\omega)$.  In this case the in-phase second harmonic response is

\begin{equation}
P = 2 |E_i|^2 sin(\beta)Im(t_{xx}^*t_{xy})
\label{PIm_theta_F}
\end{equation}

\noindent Calibrating separately $\beta(\omega)$ and the transmitted
spectral intensity, $S \simeq \frac{|E_i|^2}{2}(t_{xx}^*t_{xx})$, we find,
 
\begin{equation}
Im(\theta_F) = Im \left( \frac{t_{xy}}{t_{xx}} \right)=
\left[
\frac{P}{4 S \times sin(\beta)}
\right]
\label{Im_theta_F}
\end{equation}

Fig. \ref{Fig1} plots the directly measured Re($\theta_F$) and
Im($\theta_F$) parts of the Faraday angle.  The agreement of
Im($\theta_F$) with the Kramers-Kronig (K-K) transform of Re($\theta_F$)
confirms the consistency of our measurements, and justifies the use of the
lower noise, wider bandwidth K-K transform to represent Im($\theta_F$) in
the following calculations.  

Combining this complex $\theta_F$ with the complex longitudinal
conductivity determined from extended-Drude fits to the transmission data
\cite{Allen,Puchkov} (inset, Fig. \ref{Fig2}), the Hall angle is
determined over the full 20-250~cm$^{-1}$ range using Eq.~\ref{theta_H}
(Fig.~\ref{Fig2}). At low frequencies, we see that Re($\theta_H^{-1}$)
extrapolates to the dc transport values measured on a separate sample
using the Van der Pauw geometry, and the dc points exhibit the $T^2$ law
universally observed in the cuprates.  

We first check consistency with the linear-scattering (Lorentzian) form of
Refs. \cite{Anderson,Coleman,Lee,Ioffe,Zheleznyak,Stojkovic,Kotliar} by
solving Eq.~\ref{linscat} for the Lorentzian parameters
$\omega_H^*(\omega) = -\omega / Im[\theta_H^{-1}(\omega)]$ and
$\gamma_H^*(\omega) = \omega_H(\omega) Re[ \theta_H^{-1}(\omega)]$ in Fig.
\ref{Fig3}.  This parametrization demonstrates that $\theta_H$ is {\em
not} Lorenztian over the full frequency range since $\gamma_H(\omega)$
shows additional frequency dependence, decreasing significantly above
100~cm$^{-1}$ particularly at low temperatures.  This result is puzzling,
because ac conductivity \cite{Puchkov} and angle resolved photoemission
(ARPES) \cite{Valla} both show the generally expected {\em increase} in
scattering at higher frequencies.

For the moment, we assume this behavior may result from other energy
scales in the problem not treated by the references above.  (The
$\pi$-$\pi$ resonance, superconducting gap energy, and $\omega \sim T$ all
occur around $300-500$ cm$^{-1}$.)  Accordingly, we will focus on the
low-frequency limit to critically evaluate the Lorentzian theories. We
plot the dc limit values ${\gamma_H^*}^{dc}= \lim_{\omega\to 0}
\gamma_H^*(\omega)$ and ${\omega_H^*}^{dc} = \lim_{\omega\to 0}
\omega_H^*(\omega)$ in the inset of Fig.~\ref{Fig3} as averaged from 20 to
80 cm$^{-1}$.  At 95 K the data demonstrate consistency with previous
experiments plotted as open squares \cite{Kaplan}.  Looking at the
temperature dependence, we immediately see that ${\gamma_H^*}^{dc}(T)  
\sim T$ and ${\omega_H^*}^{dc}(T) \sim 1/T$.  This confirms the Lorentzian
behavior proposed in one spinon-holon model \cite{Lee} and in a
skew-scattering model \cite{Kotliar}.  However, this data does not support
the other Lorentzian theories which posit a temperature {\em independent}
numerator and a scattering rate quadratic in temperature
\cite{Anderson,Coleman,Ioffe,Zheleznyak,Stojkovic}. We reemphasize,
however, that all of the Lorentzian theories break down at moderate
frequecies above 100 cm$^{-1}$.

As an alternative to the linear-scattering Lorentzian model we consider a
square-scattering form for the Hall angle, which corresponds to a
Lorentzian-squared ac response.  This form was predicted by Varma and
Abrahams in their marginal Fermi liquid treatment of Hall scattering in
the cuprates \cite{Varma,Yakovenko}.  Extending their result to
finite frequencies, we get the square-Lorentzian form:

\begin{equation}
\theta_H(\omega) = \frac{\omega_H^*\Omega_p^*}
{(\Gamma_H^* - i \omega)^2}
\label{sqrscat}
\end{equation}

\noindent $\omega_H^*$ is still the cyclotron frequency, linear in $B$,
and $\Omega_p^*$ is indicative of a new energy scale in the problem,
interpreted as a Fermi surface average of the scattering derivative
\cite{Varma}.  Solving for the square-scattering rate $\Gamma_H^*$ and
weighting parameter $\omega_H^*\Omega_p^*$, one sees at once the
independence of $\omega_H^*\Omega_p^*$ from both temperature and frequency
in Fig.~\ref{Fig4}.  The existence of such a robust parameter that is
constant over 20-250~cm$^{-1}$ in infrared energy and 95-190 K in
temperature is remarkable, and provides convincing evidence that the
square-Lorentzian analysis may elucidate the dominant physics.  In the dc
limit, the scattering rate, ${\Gamma_H^*}^{dc}$, is again linear in
temperature, but with twice the value as in the Lorentzian analysis.  At
higher frequencies ${\Gamma_H^*}(\omega)$ now shows a moderate {\em
increase}, qualitatively consistent with scattering rates observed in ac
conductivity \cite{Puchkov} and ARPES \cite{Valla}.

Although the square-Lorentzian functional form for the Hall angle is a
valid causal response function, it cannot be correct at all frequencies
since it leads to a Hall sum of zero \cite{Drew}:  $\omega_H =
\frac{2}{\pi}\int^\infty_0 Re(tan [\theta_H(\omega)]) d\omega = 0$,
inconsistent with the finite $\omega_H$ implied by ARPES measurements of
the Fermi surface \cite{Valla} and band theory.  The positive Hall angle
measured at 1000~cm$^{-1}$ by Cerne et al. \cite{Cerne} is an indication
that the functional form is already changing.  The exact behavior of this
crossover is therefore an interesting topic for further investigation.

It is interesting to consider not only $\Gamma_H^*$, the renormalized ac
observable deduced from the square-Lorentzian analysis, but also the {\em
bare} scattering $\Gamma_H$.  At dc, the bare and renormalized values
satisfy $\omega_H\Omega_p/{\Gamma_H}^2 = \omega_H^*\Omega_p^* /
{\Gamma_H^*}^2$.  This bare Hall scattering $\Gamma_H$ is linear in
temperature just as the bare longitudinal scattering $\gamma_o$ is linear
in temperature as seen in the famous $\rho \sim T$ relation.  The
renormalizations enter differently, however, between longitudinal and Hall
transport since $\Gamma_H^*$ stays linear in temperature, but $\gamma_o^*$
increases superlinearly due to a decreasing renormalized transport mass
$m_o^*$ \cite{Puchkov}.  This difference might be accounted for with
vertex corrections in the Kubo formula which enter differently in
$\sigma_{xx}$ and $\sigma_{xy}$.  The $cot(\theta_H) \sim {\Gamma_H^*}^2$
form of the Hall angle obtained by ref. \cite{Varma} is itself a
consequence of vertex corrections in $\sigma_{xy}$.  Therefore, the
relation between $\gamma_o^*$ and $\Gamma_H^*$ presents an interesting
subject for future theoretical and experimental work.

In summary, we have measured the complex Hall response at infrared
frequencies.  The observed $\theta_H(\omega)$ does not fit the Lorentzian
lineshape predicted by many models of transport in the cuprates or
conventional transport theory; only at low frequencies is the data
consistent with a subset of linear-scattering models \cite{Lee,Kotliar}
that predict the observed behavior: $\gamma^*_H(T) \sim T, \omega^*_H(T)
\sim 1/T$. Alternately, the data show a good fit to a square-Lorentzian
form, over the {\em entire} frequency range, with the temperature
dependence of the response function coming again from a relaxation rate
linear in temperature, $\Gamma_H^*(T) \sim T$.  In all cases, the
experiments suggest that transport in the cuprates is governed throughout
by a relaxation rate that is linear in temperature.

{\it Acknowledgement --} The authors are grateful to E. Abrahams, J.
Cerne, A.  Millis, N. P. Ong, C. M. Varma and V. Yakovenko for insightful
discussions.  

This work was supported in part by the NSF (Grant No. DMR0070959).  M.G.
thanks the A.v. Humboldt Foundation for support during the writing of this
paper.

$*$ Current address:  {\it Walter Schottky Institut, Technische
Universit\"at M\"unchen, D-85748 Garching, Germany; email:
mgrayson@alumni.princeton.edu}

$\dag$ Current address:  {\it Laboratoire National des Champs
Magn\'etiques Puls\'es, 143 Avenue de Rangueil, 31432 Toulouse, France}

\begin{figure} 
\caption{The measured complex Faraday angle.  Insets shows separate
experimental setups for measuring Re($\theta_F$) and Im($\theta_F$).
The Kramers-Kronig transform of Re($\theta_F$) is plotted in gray.
} 
\label{Fig1}
\end{figure}

\begin{figure} 
\caption{The derived complex Hall angle.  Inset shows transmission data
with extended Drude conductivity fits for each temperature in dashed
lines.
}
\label{Fig2} 
\end{figure}

\begin{figure} 
\caption{Lorentzian parameterization of Hall angle.  The Hall scattering
$\gamma_H$ develops strong downwards frequency dependence above
100~cm$^{-1}$, implying a deviation from simple Lorentzian behavior.  
Inset plots the low-frequency limit of these parameters vs. T.
}
\label{Fig3} 
\end{figure}

\begin{figure}
\caption{Square-Lorentzian parameterization of Hall angle.  The scattering
parameter $\Gamma_H$ shows a moderate increase with frequency and
$\omega_H\Omega_p$ is notably temperature and frequency independent.  
Inset plots the low-frequency limit of these parameters vs. T.
}
\label{Fig4}
\end{figure}

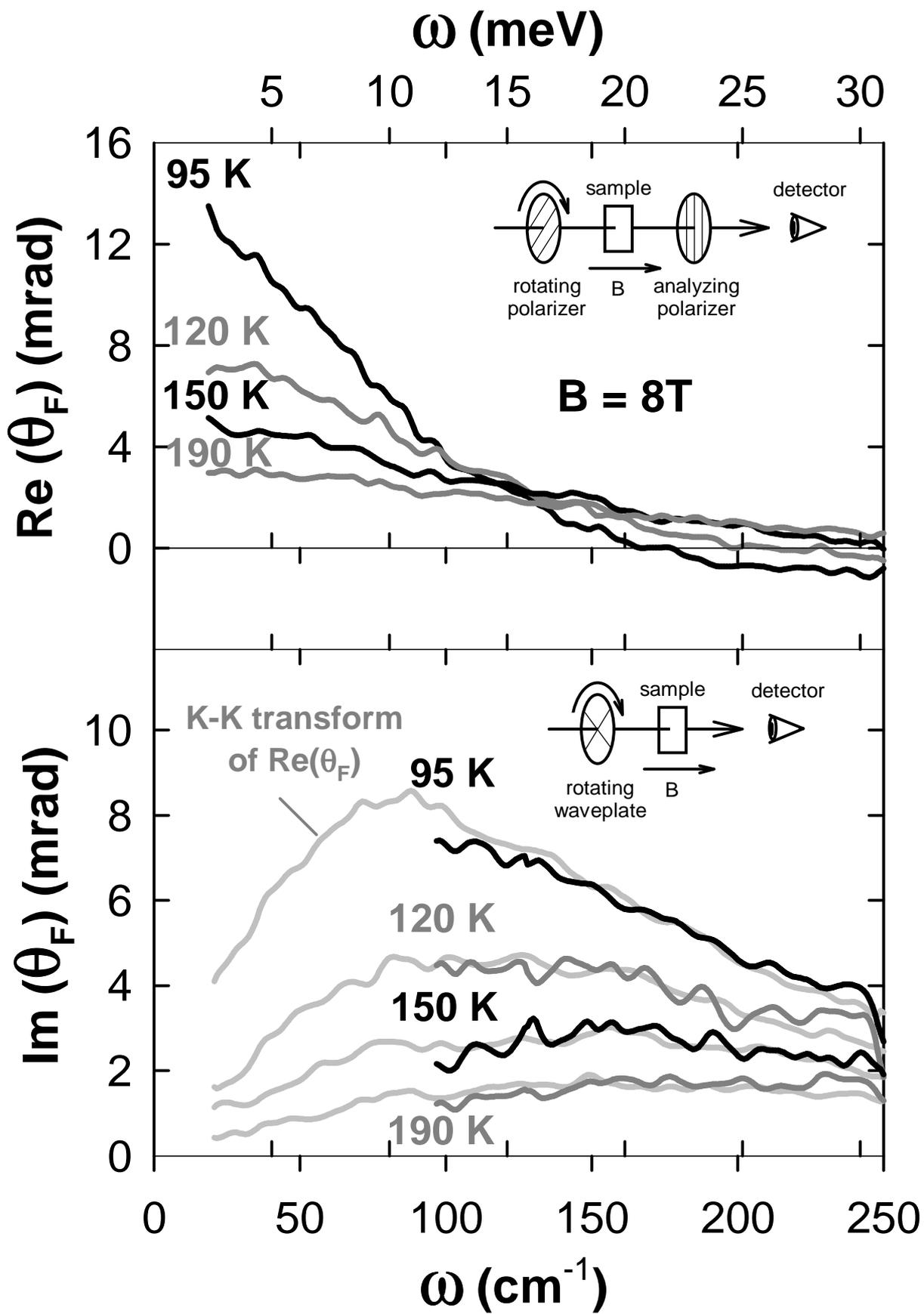
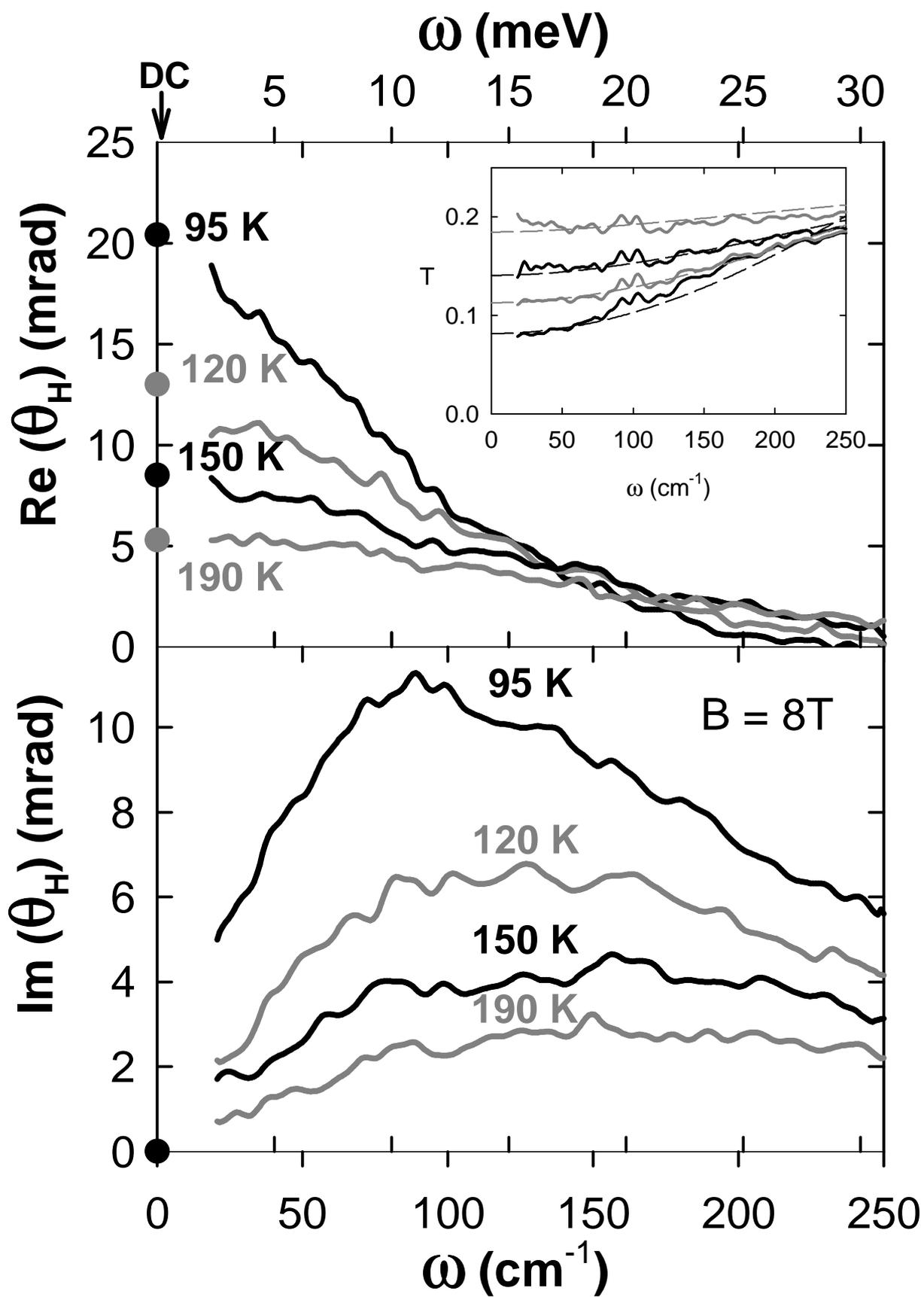
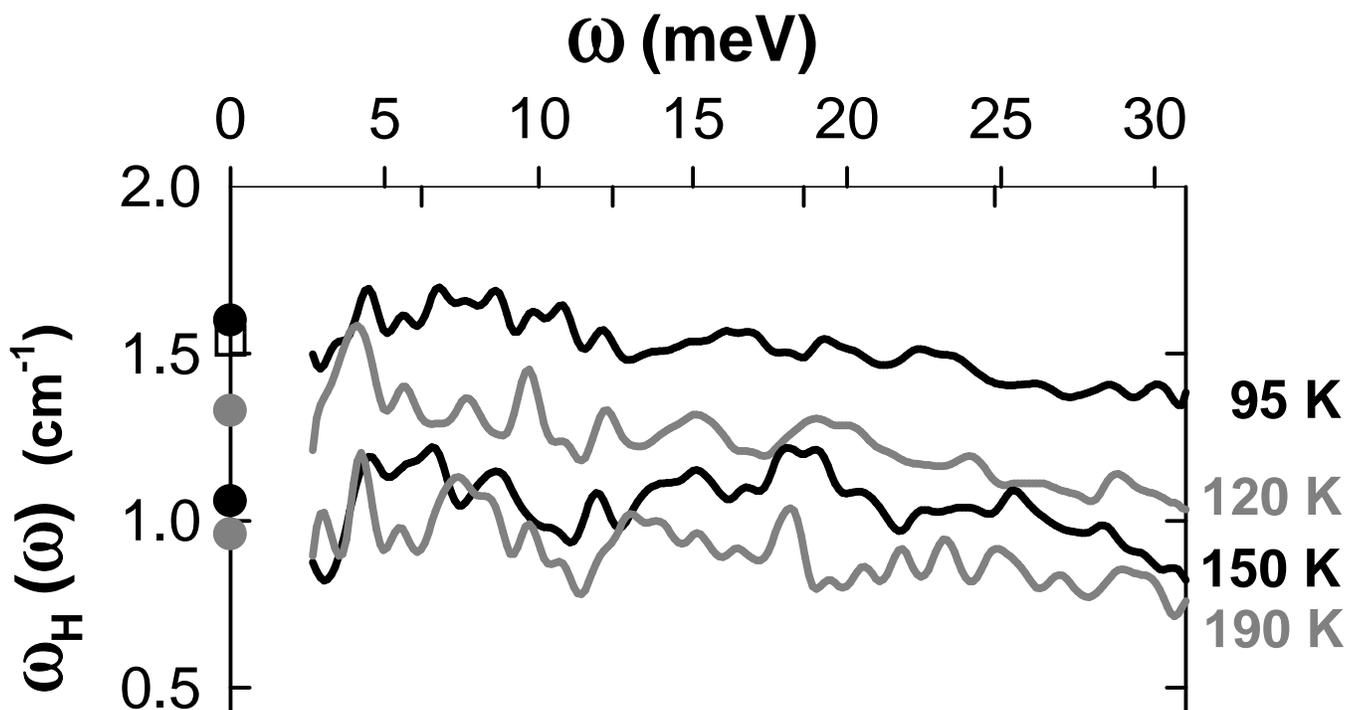
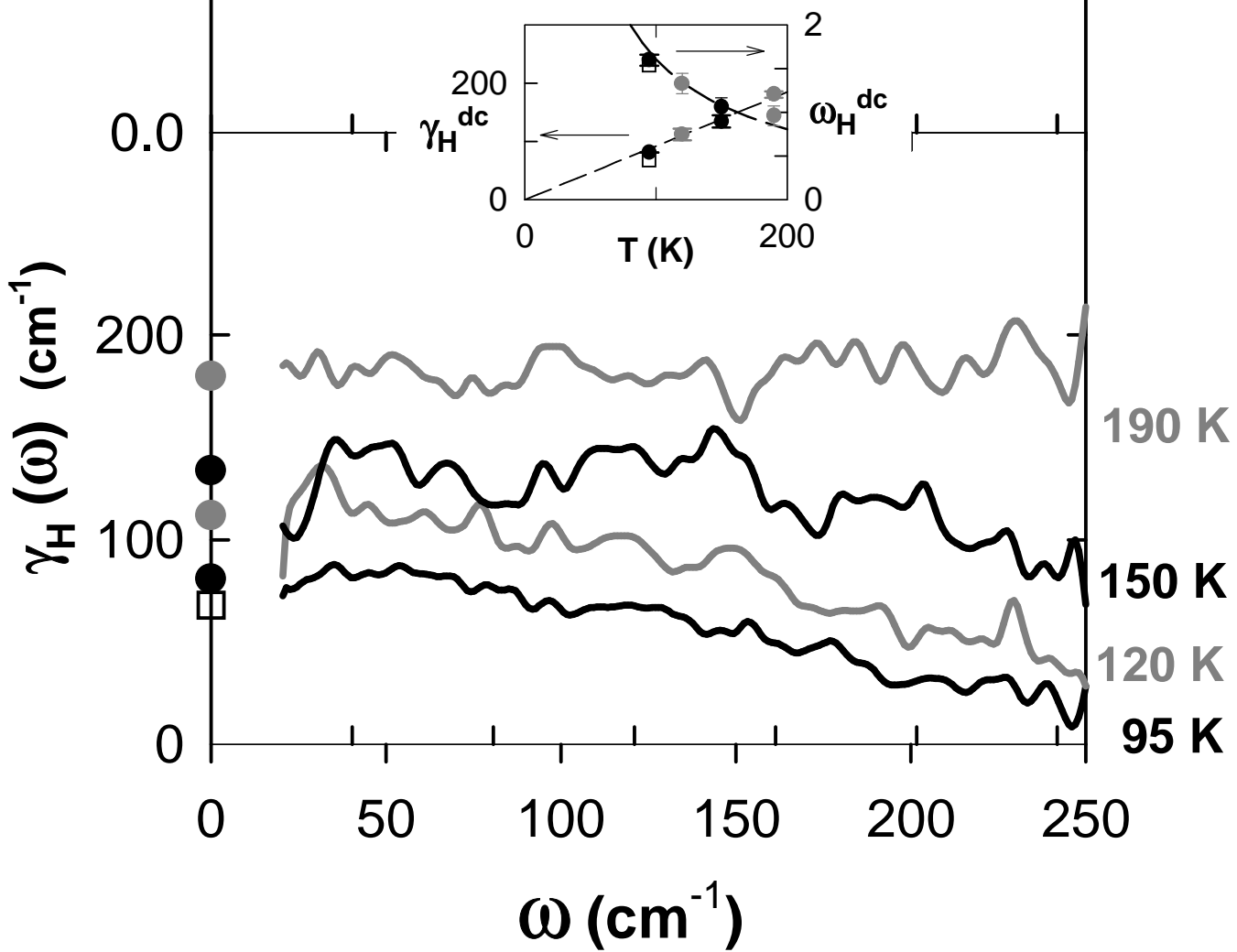
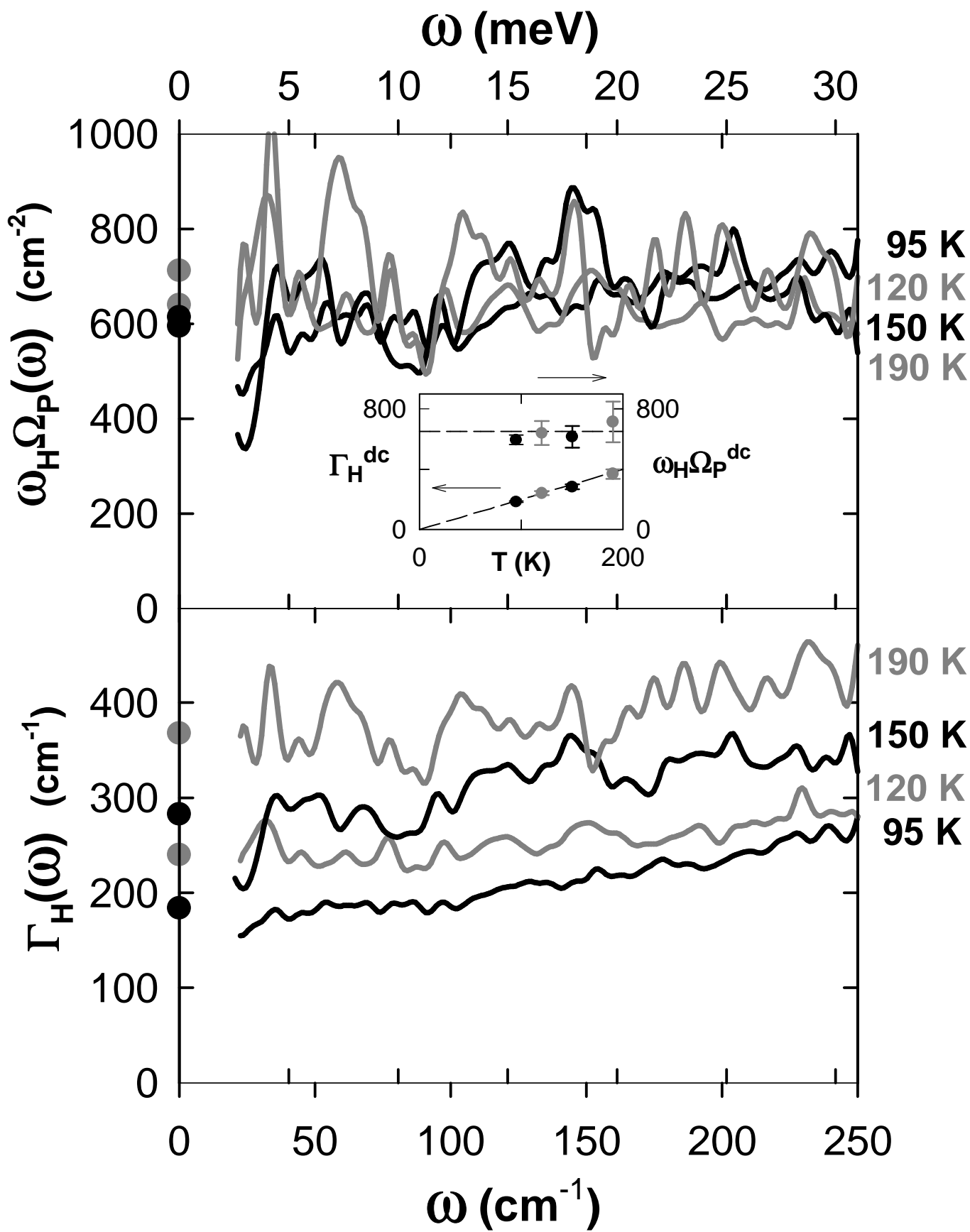
\end{document}